\documentclass[11pt,english]{article}
\usepackage[T1]{fontenc}
\usepackage[a4paper]{geometry}
\usepackage[plainpages=false, colorlinks=true, anchorcolor=blue, linkcolor=blue, citecolor=blue, bookmarks=false]{hyperref}
\usepackage{float} 
\usepackage[utf8]{inputenc}
\usepackage{amsmath}
\usepackage{amssymb}
\usepackage{babel}
\usepackage{color}
\usepackage{graphicx} 

\begin{document}

\title{Observational constraints on varying fundamental constants in a minimal CPC model}

\author{R. R. Cuzinatto$^{1,2}$\footnote{rcuzinat@uottawa.ca; rodrigo.cuzinatto@unifal-mg.edu.br}, 
R. F. L. Holanda$^{3}\footnote{holandarfl@fisica.ufrn.br}$, 
S. H. Pereira$^4,$\footnote{s.pereira@unesp.br}\\
\\
$^1$Department of Physics, \\ University of Ottawa, \\ Ottawa, ON, K1N 6N5, Canada
\\ \\
$^2$Instituto de Ci\^encia e Tecnologia, \\ Universidade Federal de Alfenas, \\ Po\c cos de Caldas, MG, 37715-400, Brazil
\\ \\
$^3$Departamento de F\'isica Te\'orica e Experimental, \\Universidade Federal do Rio Grande do Norte, \\Natal, RN, 59300-000, Brazil
\\ \\
$^4$Departamento de F\'isica, \\ Faculdade de Engenharia de Guaratinguet\'a,\\
Universidade Estadual Paulista (UNESP),  \\ Guaratinguet\'a, SP,  12516-410, Brazil
} 


\date{}

\maketitle

\begin{abstract}
A minimal model based on the Co-varying Physical Couplings (CPC) framework for gravity is proposed. The CPC framework is based on the assumptions of a metric-compatible four-dimensional Riemannian manifold where a covariantly conserved stress-energy tensor acts as source of the field equations which are formally the same as Einstein field equations, but where the couplings $\{ G, c,\Lambda \}$ are allowed to vary simultaneously. The minimal CPC model takes $\Lambda$ as a genuine constant while $c$ and $G$ vary in an entangled way that is consistent with Bianchi identity and the aforementioned assumptions. The model is constrained using the most recent galaxy cluster gas mass fraction observational data. Our result indicates that the functions $c(z)$ and $G\left(z\right)=G_{0}\left(c/c_{0}\right)^{4}$ are compatible with constant couplings for the three different parameterizations of $c=c(z)$ adopted here.
\end{abstract}


\newpage

\tableofcontents{}


\section{Introduction \label{sec-Intro}}

The flat $\Lambda$CDM model of cosmology, known as cosmic concordance model, correctly predicts the overall evolution of the Universe, including the current phase of acceleration. It is also consistent with many cosmological and astrophysical observations, which makes general relativity (GR) the protagonist as the fundamental theory of gravity. However, even though it is successful at different scales and explains very diverse data sets, there exist some specific observational discrepancies and even theoretical shortcomings that open up the possibility for extensions of GR. As few examples, we mention the recent observation that the Universe is expanding faster than expected  \cite{Riess:2019qba}, the cosmic curvature problem \cite{DiValentino:2019qzk,Handley:2019tkm}, small-scale problems \cite{Peebles:2021gou} and the statistically significant tension between the predictions for the Hubble constant value by early-time probes assuming the $\Lambda$CDM model and the corresponding determinations of $H_0$ by a number of late-time model-independent fits from local measurements; this incongruity is  the so called Hubble tension \cite{Martinelli:2019krf,DiValentino:2021izs}.  For an interesting recent review on the $\Lambda$CDM model problems, we refer the reader to Ref.~\cite{Bull:2015stt}. 

The above-mentioned problems of the concordance cosmological model expose some of GR's shortcomings in the infrared regime; on top of that, the ultraviolet regime of gravity also hints limitation of its description by GR. Within the cosmological context, the early universe puzzles demanded the introduction of the inflation paradigm \cite{Brandenberger:1999sw,Mukhanov:2005sc,Baumann:2009ds}. Moreover, very-high energy/curvature regimes would demand a quantum version of GR that is not currently available in a final consistent form---in spite of the loop quantum gravity \cite{Ashtekar:2021kfp} and string theory \cite{Kiritsis:2019npv} proposals. The early works by Stelle and others were an attempt to alleviate the quantization issues of GR via the addition of $R^2$ \cite{Stelle:1976gc,Stelle:1977ry} and other higher-order curvature terms to the Einstein-Hilbert action \cite{Lee:1970iw}. Starobinsky model also makes use of the $R^2$ term to realise an inflationary scenario derived from an action principle beyond that of GR \cite{Gurovich:1979xg,Starobinsky:1979ty,Starobinsky:1980te}. From a more theoretical standpoint, one could also argue that the scope of GR is somewhat limited since it is based on a Riemannian manifold that is metric compatible and torsion free; as such it does not include teleparallel gravity \cite{Aldrovandi:2013wha,Bahamonde:2021gfp}, Riemann-Cartan gravity \cite{DeSabbata:1986sv,Minkevich:2016wya} or metric-affine gravity \cite{Iosifidis:2019dua}. Moreover, GR is not a typical gauge theory since it exhibits universality and soldering properties \cite{Aldrovandi:1996ke}. These limitations of GR and its applications to cosmology may serve as motivation for pursuing modified gravity theories.

Some examples of modified gravity theories are Horndeski \cite{Horndeski:1974wa,Charmousis:2011bf,Kobayashi:2019hrl},
DHOST \cite{Langlois:2015cwa,BenAchour:2016cay,Langlois:2018dxi}, and $f(R)$
theories of gravity \cite{sotiriou2010f,DeFelice:2010aj,Nojiri:2017ncd}. There is also
the class of $f\left(R,\nabla R,\dots,\nabla^{n}R\right)$ theories
\cite{cuzinatto2016scalar} which can be viewed as effective field theories
stemming from a more complete and fundamental theory in certain energy/curvature regimes; the $f\left(R,\dots,\nabla^{n}R\right)$ proposal has applications in gauge theory of gravity \cite{Cuzinatto:2006hb}, black-hole physics \cite{Cuzinatto:2017srn}, gravitational waves \cite{Vilhena:2021bsx}, inflation
\cite{Cuzinatto:2018vjt,Cuzinatto:2018chu}, and dark-energy \cite{Cuzinatto:2013pva}. Both $f\left(R\right)$ and $f\left(R,\dots,\nabla^{n}R\right)$ theories
were shown to be equivalent to Brans-Dicke (BD) theory \cite{Brans:1961sx}
for specific values of its characteristic dimensionless parameter
$\omega$ \cite{sotiriou2010f,cuzinatto2016scalar}. In fact, Brans-Dicke
theory is the paradigm for a large class of alternative proposals
to GR \cite{Faraoni:2004pi}. BD follows from other frameworks too, such as a result of Kaluza-Klein after compactification \cite{Kaluza:1921tu,Klein:1926tv,Bailin:1987jd,Overduin:1997sri}. BD can be derived as a limit of string theory: the lower-energy limit of bosonic string leads to a BD model
with $\omega=-1$ \cite{Callan:1985ia,Fradkin:1985ys}. According to 
\cite{Faraoni:2004pi}, BD theory is accommodated within Lyra's geometry \cite{lyra1951modifikation,Sen:1971fac}---see also \cite{Cuzinatto:2021ttc}.

BD theory, its scalar-tensor generalizations---see e.g. \cite{Ballardini:2020iws,Braglia:2020auw,Ballardini:2021evv}---, and other extensions of the current standard model naturally lead to the concept that fundamental couplings are indeed expected to vary. Some examples of varying physical couplings proposals are studies of variation of the fine-structure constant $\alpha$ \cite{King:2012id,Galli:2012bf,Leefer:2013waa,vandeBruck:2015rma,Kotus:2016xxb,Colaco:2019fvl,Goncalves:2019xtc,Liu:2021mfk,Bora:2020sws}, of the Newton's gravitational constant $G$ \cite{dirac1937cosmological,Jofre:2006ug,Verbiest:2008gy,Lazaridis:2009kq,Garcia-Berro:2011kvq,Ooba:2016slp,Zhao:2018gwk,Vijaykumar:2020nzc}, and of the speed of light $c$ \cite{FermiGBMLAT:2009nfe,Cruz:2012bwp,Salzano:2014lra,Qi:2014zja,Xu:2016zxi,Xu:2016zsa,Cao:2016dgw,Cao:2018rzc,Salzano:2016hce,Liu:2018qrg,Liu:2021eit,Agrawal:2021cim,Mendonca:2021eux,Zhu:2021pml}. The common feature of these works is to take only one of the couplings 
as a function of the cosmological time while disregarding possible variations of the others. Variation of more than one of the aforementioned couplings is a possibility that has been studied recently by \cite{Franzmann:2017nsc,Costa:2017abc,Gupta:2020anq,gupta2021constraint,Gupta:2021wib,Gupta:2020wgz,Gupta:2021eyi,Gupta:2021tma,Gupta:2022amf,Bonanno:2020qfu,Lee:2020zts,Cuzinatto:2022vvy}, which include the cosmological constant $\Lambda$---among others---to the set. More specifically, Refs. \cite{Franzmann:2017nsc,Costa:2017abc,Gupta:2020anq} consider the simultaneous variations of $\left\{ c,G,\Lambda\right\}$ within the so called Co-varying Physical Couplings (CPC) framework. For example, Ref. \cite{Costa:2017abc} introduces a covariant $c$-flation mechanism to resolve the
early-universe puzzles usually addressed by inflation.
Ref. \cite{Gupta:2020anq} applies the CPC scheme to the late-time universe
for a particular ansatz for the time-varying trio $\left\{ c,G,\Lambda\right\} $. The entangled variation of the couplings is a crucial feature of the CPC framework, one that distinguishes it from earlier approaches that account for varying physical constants, and a characteristic whose importance was already pointed out in Refs. \cite{Petit:1995ys,Barrow:1999is}.

The Varying Speed of Light (VSL) proposals are examples of scenarios accounting for the variation of $c$ alone---see e.g. \cite{Moffat:1992ud,Albrecht:1998ir}, and \cite{Uzan:2002vq} for a review. The use of Baryon Acoustic Oscillations
(BAO) to constrain the value of $c$ was worked out recently by \cite{Salzano:2014lra}, based on a relation between the Hubble parameter function and the angular diameter distance at the same redshift. In \cite{Cao:2016dgw,Salzano:2016hce}, the measurement of $c$ was done by utilising radio quasars calibrated as standard rulers.  Ref.  \cite{Qi:2014zja} used type Ia Supernovae (SNe Ia), BAO, $H(z)$, and CMB data to constrain the variation of $c$ as $\sim 10^{-2}$, a result confirmed by \cite{Liu:2021eit} through a combination of SNe Ia and strong gravitational lensing (SGL). In \cite{Mendonca:2021eux} a new method to test the eventual dependence of the speed of light on the redshift was implemented by combining the measurements of galaxy cluster gas mass fraction (GMF), $H(z)$ and SNe Ia; the analyses  indicate negligible variation of the speed of light.
 
In this paper, we will advance in the proposal of studying the eventual variation of more than one fundamental constant simultaneously, namely, we shall investigate possible variations of $\{c,G\}$. Accordingly, we will perform a consistency check for the time-independence of the speed of light $c$ and of the gravitational coupling $G$ in the context of the Co-varying Physical Couplings framework; this will be done by constraining the minimal CPC model with the most recent GMF data obtained from galaxy clusters \cite{SPT:2021vsu} and complementary data from  SNe Ia \cite{Pan-STARRS1:2017jku}. The minimal CPC model is a particular realization of the CPC scheme that neglects possible variations of $\Lambda$ and entangles the eventual variations of $c(z)$ and $G(z)$ in a consistent way, viz. $G\left(z\right)=G_{0}\left(c/c_{0}\right)^{4}$. {Three} different parametrizations for $c(z)$ are analysed.
 
The paper is organized as follows: Section \ref{sec-TheoreticalBackground} contains the theoretical background focused on Brans-Dicke theory of gravity as an inspiration for the Co-varying Physical Couplings framework. The CPC scenario is built in Section \ref{sec-CPCframework} where the general constraint relating the simultaneous variation of $\{G,c,\Lambda\}$ is derived. In Section \ref{sec-minimalCPCmodel} we choose the minimal realization of the CPC scenario described in the end of the previous paragraph. Section \ref{sec-DataFitting} {and the following deal with} the fitting of the minimal CPC model to the galaxy cluster gas mass fraction data. Section \ref{sec-GMF} explains the data and how the related equations are adapted in the face of varying couplings; and, in Section \ref{sec-GMFconstrainsCPC}, the plots and numerical values of the free parameters of the model are presented and discussed. Our conclusions and final comments are stated in Section \ref{sec-Conclusion}.


\section{Theoretical background \label{sec-TheoreticalBackground}}

BD equations of motion are given by \cite{Faraoni:2004pi}:
\begin{align}
R_{\mu\nu}-\frac{1}{2}g_{\mu\nu}R+\frac{V}{2\phi}g_{\mu\nu} & =\frac{8\pi}{\phi}T_{\mu\nu}+\frac{\omega}{\phi^{2}}\left(\nabla_{\mu}\phi\nabla_{\nu}\phi-\frac{1}{2}g_{\mu\nu}\nabla^{\rho}\phi\nabla_{\rho}\phi\right)\nonumber \\
 & \hphantom{=\frac{8\pi}{\phi}T_{\mu\nu}\;}+\frac{1}{\phi}\left(\nabla_{\mu}\nabla_{\nu}\phi-g_{\mu\nu}\square\phi\right),\label{eq:BD-FE}
\end{align}
where $g_{\mu\nu}$ is the space-time metric, $R_{\mu\nu}$ is the
Ricci tensor, $R$ is the curvature scalar, $T_{\mu\nu}$ is the matter energy momentum tensor, $\phi$ is BD scalar field and $V$ is the associated potential. Differently from GR, BD equations of motion fully incorporate Mach's principle by realising the idea that the gravitational coupling $G$ is an effect due to the whole matter in the universe. In fact, comparison of (\ref{eq:BD-FE}) with Einstein's equations of GR suggests that BD scalar field $\phi$ plays the hole
of the inverse gravitational coupling:
\begin{equation}
G=\frac{1}{\phi}.\label{eq:BD-phi(G)}
\end{equation}
BD equation (\ref{eq:BD-FE}) is supplemented by the equation of motion
for $\phi$, namely:
\begin{equation}
\frac{2\omega}{\phi}\square\phi+R-\frac{\omega}{\phi^{2}}\nabla^{\rho}\phi\nabla_{\rho}\phi-\frac{dV}{d\phi}=0\label{eq:BD-phi-EOM}.
\end{equation}
Therefore, $\phi$ is a dynamical field. Due to (\ref{eq:BD-phi(G)}),
a dynamical scalar field $\phi=\phi\left(x^{\mu}\right)$ implies
a $G$ that is a function of the space-time coordinates. In this way,
BD theory admits a varying physical coupling, viz. Newton's gravitational
constant. 

Notice that BD theory reduces to standard GR in the limit of $\omega\rightarrow\infty$
\cite{Weinberg:1972kfs}. Solar system tests require large values of $\omega$,
typically $\omega\gtrsim40,000$ \cite{Will:2018bme}. This means that pure BD should be very close to GR in local scales. This is a shortcoming
that could be potentially eliminated by a screening mechanism \cite{Khoury:2003rn,Ferreira:2018wup}.
Otherwise, one could argue that an extension of BD is of interest
in cosmological scales. Our proposal aligns with the latter possibility.


\subsection{The CPC framework \label{sec-CPCframework}}

BD theory yields a varying gravitational coupling, $G=G\left(x^{\mu}\right)$.
This naturally incepts the idea to relax the constancy of the other
couplings appearing in the field equations for the gravitational field.
In the Einstein field equations of GR, these couplings are the speed of
light $c$, the gravitational coupling $G$, and the cosmological
term $\Lambda$. Accordingly, we promote these couplings to space-time functions: $c=c\left(x^{\mu}\right)$, $G=G\left(x^{\mu}\right)$, and
$\Lambda=\Lambda\left(x^{\mu}\right)$. For reasons that will be clear
below, we call this scenario the Co-varying Physical Couplings (CPC)
framework.

Refs. \cite{Franzmann:2017nsc,Costa:2017abc} have proposed a setup that
is a minimal generalization of GR inspired by the BD feature of a
varying $G$ from the following requirements: 
\begin{enumerate}
\item The framework for gravity is covariant. The field equations are \emph{tensorial}
in character; they are preserved under general coordinate transformations;
they are valid on a \emph{metric-compatible} (Riemannian) four-dimensional
manifold \cite{Weinberg:1972kfs}: 
\begin{equation}
\nabla_{\rho}g_{\mu\nu}=0.\label{eq:metricity}
\end{equation}
The covariant derivative $\nabla_{\rho} = \partial_{\rho} + \Gamma_{\rho}$ involves the Christoffel symbols, a connection that is typical from GR; this connection is torsion free, and it admits a curvature tensor.

\item The energy momentum tensor is (covariantly) conserved \cite{Carroll:2004st}:
\begin{equation}
\nabla_{\mu}T^{\mu\nu}=0.\label{eq:Tmunu-conservation}
\end{equation}
There is a number of VSL proposals violating this basic requirement,
e.g. the VSL model in \cite{Albrecht:1998ir}.

\item The equation of motion for the gravitational field is \emph{formally}
the same as\textbf{ }Einstein Field Equations (EFE)\textbf{ }of GR.
We want to produce minimal changes to GR, therefore EFE are maintained except
that the couplings $\left\{ G,c,\Lambda\right\} $ \emph{are allowed
to change} as functions of the space-time coordinates. In particular,
the equation of motion for a minimal CPC will exhibit the Einstein tensor 
\begin{equation}
G_{\mu\nu}=R_{\mu\nu}-\frac{1}{2}Rg_{\mu\nu},\label{eq:Gmunu}
\end{equation}
which is (covariantly) conserved: 
\begin{equation}
\nabla_{\mu}G^{\mu\nu}=0.\label{eq:Bianchi}
\end{equation}
This is the Bianchi identity.
\end{enumerate}

By construction, the equation of motion of CPC scenery will be the EFE-inspired equation (Feature
\#3 above):
\begin{equation}
R_{\mu\nu}-\frac{1}{2}g_{\mu\nu}R+\Lambda\left(x^{\mu}\right)g_{\mu\nu}=\frac{8\pi G\left(x^{\mu}\right)}{c^{4}\left(x^{\mu}\right)}T_{\mu\nu}.\label{eq:CPC-FE}
\end{equation}
Our intention is to depart from GR as little as possible. By keeping
the form of Einstein Field Equations, we guarantee that the 
Co-varying Physical Couplings scenario recovers standard GR in the
traditional case where $c$, $G$, and $\Lambda$ are genuine constants. 

Taking the covariant derivative 
of Eq. (\ref{eq:CPC-FE}) leads to:
\begin{equation}
\nabla^{\mu}G_{\mu\nu}+\left(\nabla^{\mu}\Lambda\right)g_{\mu\nu}+\Lambda\left(\nabla^{\mu}g_{\mu\nu}\right)=\nabla^{\mu}\left(\frac{8\pi G}{c^{4}}\right)T_{\mu\nu}+\frac{8\pi G}{c^{4}}\left(\nabla^{\mu}T_{\mu\nu}\right).\label{eq:nabla-CPC-EF}
\end{equation}
Now, due to Eqs. (\ref{eq:metricity}), (\ref{eq:Tmunu-conservation}),
and (\ref{eq:Bianchi}), several terms vanish in (\ref{eq:nabla-CPC-EF})
and we are left with
\begin{equation}
\left[\frac{\partial_{\mu}G}{G}-4\frac{\partial_{\mu}c}{c}\right]\frac{8\pi G}{c^{4}}T^{\mu\nu}-\left(\partial_{\mu}\Lambda\right)g^{\mu\nu}=0,\label{eq:GC}
\end{equation}
Following \cite{Costa:2017abc}, we call this equation the General Constraint (GC). Eq.~(\ref{eq:GC}) must be satisfied by any consistent CPC model. 

A common criticism on varying physical constants proposals is their supposedly lack of an associated action principle \cite{Ellis:2003pw}. However, Ref.~\cite{Costa:2017abc} shows that this is not the case for the CPC framework. Therein, an action integral is introduced for the CPC scenario and it is demonstrated that the GC is compatible with a variational approach. 

In a quite different approach, Ref.~\cite{Bonanno:2020qfu} studies gravitational equations in the presence of matter with spacetime dependent $G$ and $\Lambda$ by making use of the Bianchi identities. The authors of this paper obtain a condition resembling our Eq.~(\ref{eq:GC}), although they do not admit a varying $c$. Moreover, the formalism developed by \cite{Bonanno:2020qfu} is applied in the context of the Asymptotic Safety scenario to the early universe. Herein, we will apply a minimal configuration of CPC framework to the late-time universe.


\subsection{The minimal CPC model \label{sec-minimalCPCmodel}}

The consequences of the GC in Eq.~(\ref{eq:GC}) are striking. First, one notices that the energy momentum tensor multiplies the
square brackets in the GC; therefore, the eventual variations of $\left\{ c,G,\Lambda\right\} $
are coupled to the matter fields via $T_{\mu\nu}$. Hence, our minimal
generalization of BD theory that keeps the form of Einstein field
equations modifies the traditional saying by GR to: ``Matter ($T_{\mu\nu}$)
tells space-time how to curve ($G_{\mu\nu}$) and the couplings $\left\{ c,G,\Lambda\right\} $ how to vary; space-time and the varying couplings tell matter how to
move''.

Second, the eventual variations of $\left\{ c,G,\Lambda\right\} $
are not independent. If one of the quantities $G$, $c$ or $\Lambda$
vary, then the other(s) must also vary for the sake of internal consistency
of the framework via the GC in Eq.~(\ref{eq:GC}). This is true even
in the simple model where one of the three quantities $\left\{ c,G,\Lambda\right\} $ is constant. Indeed, take the case where
\begin{equation}
\Lambda=\text{constant},\label{eq:Lambda_constant}
\end{equation}
for instance. Plugging (\ref{eq:Lambda_constant}) into (\ref{eq:GC}):
\begin{equation}
\begin{cases}
\partial_{\mu}\Lambda=0\\
T^{\mu\nu}\neq0
\end{cases}\Rightarrow\left[\frac{\partial_{\mu}G}{G}-4\frac{\partial_{\mu}c}{c}\right]=0.\label{eq:GC(c,G)}
\end{equation}
Notice that vacuum solutions were disregarded, i.e. we have taken
$T^{\mu\nu}\neq0$. This is the realistic case of the present-day
universe, which is filled with radiation and matter (baryonic and
dark) besides the cosmological constant $\Lambda$. The solution to (\ref{eq:GC(c,G)})
is:
\begin{equation}
\frac{G}{G_{0}}=\left(\frac{c}{c_{0}}\right)^{4}\qquad\left(\Lambda=\text{constant}\right),\label{eq:G(c)}
\end{equation}
where $c_{0}$ and $G_{0}$ are the values of the speed of light and
of the gravitational coupling at our current space-time coordinate.

Now the name of our framework, Co-varying Physical Couplings, should
be clear: the couplings $c$, $G$, and $\Lambda$ cannot vary independently;
they co-vary according to Eq.~(\ref{eq:GC}). Even in the particular
case described by Eq.~(\ref{eq:G(c)}), if the speed of light $c$
is allowed to vary, not only the gravitational coupling $G$ must
vary, but it must vary as $G\propto c^{4}$ according to what is dictated
by the GC.\footnote{As an interesting aside, notice that if one demands both $c$ and
$G$ to be genuine constants, then the GC enforces the cosmological
coupling $\Lambda$ to be a constant too. Actually, Eq.~(\ref{eq:GC})
demands that all the quantities in $\left\{ c,G,\Lambda\right\} $
will be necessarily constants whenever two of the quantities in the
set are taken to be genuine constants (under $T^{\mu\nu}\neq0$).} 

Here we are going to realise the minimal CPC model in Eq.~(\ref{eq:G(c)})
and test it against observational data. Although Eq.~(\ref{eq:Lambda_constant})
determines $\Lambda$ (in terms of its present-day value) and Eq.
~(\ref{eq:G(c)}) entangles the pair $\left\{ c,G\right\} $, we still
need to introduce a specific functional form of $c$
(or $G$) in terms of $x^{\mu}$. This is analogous to what happens
in standard cosmology, where we still need an equation of state to 
complement Friedmann equations. Our first supposition is that in a homogeneous and isotropic universe,
one expects the scalar functions $\left\{ c,G,\Lambda\right\} $ to
be, at most, function of the time $t$ (but not of the spacial coordinates
$\left\{ x,y,z\right\} $). Then, we will assume:
\begin{equation}
c=c\left(t\right)\,,\quad G\left(t\right)\propto c^{4}\,,\quad\text{and}\quad\Lambda=\text{constant}.\label{eq:minimal-CPC-model}
\end{equation}
Next, we propose the following parameterizations for $c(z)$: 
\begin{align}
& c\left(z\right)=c_{0}\phi_{c}\left(z\right)\,, \nonumber \\
& \phi_{c}\left(z\right)\equiv \begin{cases}
1 + c_{1} z & \quad \left(\text{Parameterization I}\right)\\
1 + c_{1} \frac{z}{\left(1+z\right)} & \quad \left(\text{Parameterization II}\right) \\
1 + c_{1} \log\left(1+z\right) & \quad \left(\text{Parameterization III}\right)
\end{cases}.
\label{eq:c(z)}
\end{align}
Eq.~(\ref{eq:c(z)}) carries the redshift $z$, defined in terms of the scale factor as $\left(1+z\right)=\left(a_0/a\right)$;
it relates with the cosmic time $t$ since $a=a\left(t\right)$. As usual, $a_0=a(t_0)$ with $t_0$ representing the present-day time. The dimensionless parameters $c_{1}$ is a measure of the departure of $c$ from the
constant value $c_{0}=299\,792\,458\,\text{m/s}$. 

The motif for {Parameterizations I and II} in (\ref{eq:c(z)}) is two-fold, namely: (i) simplicity, since that $\phi_c$ is a linear function of $z$ or $a$; and, (ii) one would expect that the speed of light (and consequently $G$) would vary slowly nowadays. Traditional VSL models do assume abrupt phase transitions for the speed of light, phases during which $c=c\left(t\right)$ changes vertiginously; however, these transitions are typically assumed to take place in the very-early universe \cite{Albrecht:1998ir}, not at the present. {Parameterization III was included for completeness: an eventual departure from linearity (in terms of $z$) could be captured by this ansatz. Moreover, the parameterization in terms of $\log$ is typical of models applied to describing the late-time universe \cite{Mamon:2016dlv,Mamon:2017rri,AlMamon:2018uby}.}

The observational window to be used for constraining the model is
the galaxy cluster gas mass fraction measurements. This probe is preferred
because it is model agnostic. That is a convenient characteristic
since we do not have to worry about how the varying $c$ and $G$
could modify the calibration used in the raw data. The next section
details the procedure of constraining the parameter $c_{1}$.


\section{Gas mass fraction measurements with varying $c$ and varying $G$ \label{sec-DataFitting}}

A quantity of interest for imposing limits on possible $G$ and $c$ variations is the so-called gas mass fraction of galaxy clusters, defined by $f_{gas}=M_{gas}/M_{tot}$, where $M_{gas}$ is the mass of the intracluster gas and $M_{tot}$ is the total mass, which includes baryonic gas mass and dark matter mass. As it is largely known, the physical process dominating in the galaxy cluster  baryonic content  is  the intracluster gas X-ray emission, detected predominantly via thermal bremmsstrahlung~\cite{Sarazin:1986zz}.

Within a radius $R$, the gas mass $ M_{gas} (<R) $ can be obtained by X-ray observations, being  written as \cite{Sarazin:1986zz}:
\begin{eqnarray}
M_{gas} (<R) &=& \left( \frac{3 \pi \hbar m_e c^2}{2 (1+X) e^6}
\right)^{1/2}  \left( \frac{3 m_e c^2}{2 \pi k_B T_e} \right)^{1/4}
m_H \nonumber\\
& &  \times \frac{1}{[\overline{g_B}(T_e)]^{1/2}}
{r_c}^{3/2} \left
[ \frac{I_M (R/r_c, \beta)}{I_L^{1/2} (R/r_c, \beta)} \right] [L_X
(<R)]^{1/2}\;,
\label{M_gas}
\end{eqnarray}
where $T_e$ is the gas temperature, $m_e$ and $m_H$ are the electron and hydrogen masses, respectively, $X$ is the hydrogen mass fraction,  $\overline{g_B}(T_e)$ is the Gaunt factor, $L_X
(<R)$ is the bolometric luminosity, $r_c$ stands for the core radius and
$$
I_M (y, \beta) \equiv \int_0^y (1+x^2)^{-3 \beta/2} x^2 dx\;,
$$
$$
I_L (y, \beta) \equiv \int_0^y (1+x^2)^{-3 \beta} x^2 dx\;.
$$
Then, from Eq.~(\ref{M_gas}),
\begin{equation}
M_{gas} (< R) \propto c^{3/2}.
\label{M_gas(c)}
\end{equation}

On the other hand, the galaxy cluster total mass  within a given radius $R$ is calculated by considering that the intracluster gas is in hydrostatic equilibrium~\cite{Sarazin:1986zz,Allen:2007ue,Allen:2011zs}, i.e.
\begin{equation}
M_{tot} (<R) = - \left. \frac{k_B T_e R}{G \mu m_H} \frac{d \ln
n_e(r)}{d \ln r} \right|_{r=R},
\end{equation}
where $n_e$ is the electron number density and $\mu$ is the mean molecular weight \cite{Sarazin:1986zz}. 
It can be seen from the above equation that
\begin{equation}
M_{tot} \propto G^{-1}.
\label{M_gas(G)}
\end{equation}
From (\ref{M_gas(c)}) and (\ref{M_gas(G)}), one concludes that the gas mass fraction $f_{gas}=M_{gas}/M_{tot}$ scales as:
\begin{equation}
\label{final}
f_{gas} \propto c^{3/2}(z) \, G(z).
\end{equation}
Finally, by using Eqs.~(\ref{eq:minimal-CPC-model}) and  (\ref{eq:c(z)}), the expression of $f_{gas}$ is cast in an appropriate form for the minimal CPC model: \begin{equation}
\label{final}
f_{gas} \equiv \phi_{c}^{11/2}(z) \frac{M_{gas}}{M_{tot}}.
\end{equation}
It is important to emphasize here that, due to Eqs. \eqref{eq:G(c)} and \eqref{eq:c(z)}, the function $\phi_c(z)$ is dimensionless. This is consistent with the main arguments found in Refs.~\cite{Ellis:2003pw,Duff:2001ba} conditioning the physical meaningfulness of theories with varying constants to the use of their corresponding dimensionless ratios.


\begin{figure*}
    \centering
    \includegraphics[width=7cm, height=6cm]{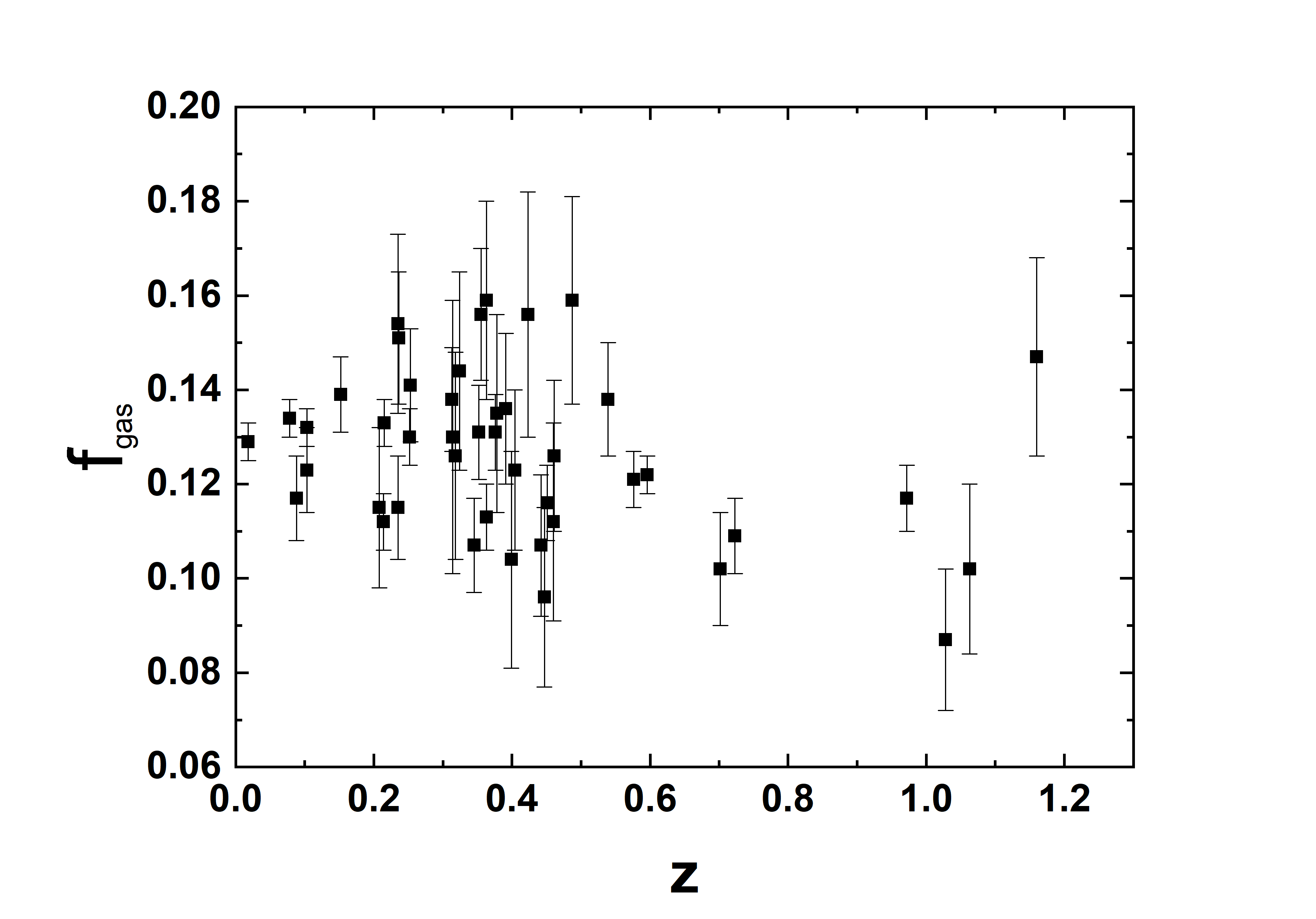} 
      \includegraphics[width=7cm, height=6cm]{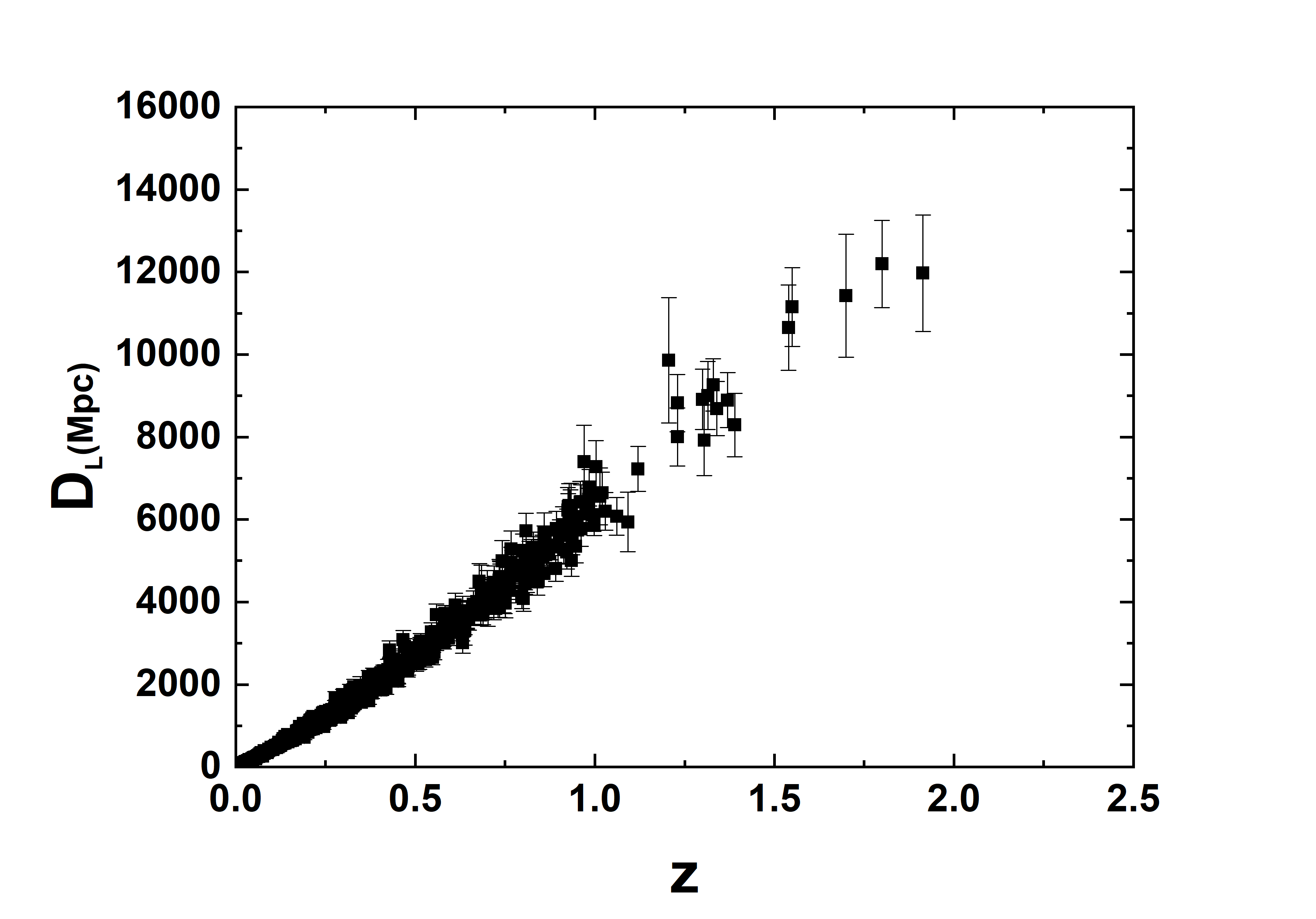} 
    \caption{Left: Gas mass fraction data. Right: the complete Pantheon SNe Ia Data.}
    \label{fig:riess1}
\end{figure*}


\section{Gas mass fraction data \label{sec-GMF}}

We shall put limits on the parameter $c_1$ in Eq.~(\ref{eq:c(z)}) by using the most recent X-ray gas mass fraction measurements of hot ($kT \geq  5$ keV), massive and  relaxed  galaxy clusters from  Chandra archive. The sample consists of 44 galaxy clusters in redshift range  $0.018 \leq z \leq 1.160$~\cite{SPT:2021vsu} (see Fig.~\ref{fig:riess1}--left).  The bias in the mass measurements from X-ray data arising by assuming hydrostatic equilibrium was calibrated using robust mass estimates for the target clusters from weak gravitational lensing \cite{Applegate:2015kua}. Moreover, the  authors of Ref.~\cite{SPT:2021vsu} measured the gas mass fraction in spherical shells at radii near $r_{2500}$, instead of using  the cumulative fraction integrated over all radii ($< r_{2500}$) as in previous works. This procedure reduces significantly the systematic uncertainties, cf. detailed discussion by \cite{Mantz:2014xba}.

On the other hand, the expected constancy of the $f_{gas}$ within massive, hot and relaxed galaxy clusters 
with redshifts in the interval of interest
has been used to constrain cosmological parameters via the following equation:
\begin{equation}
\label{fgas1}
f_{gas}(z) = \gamma(z)K(z) A(z) \left[\frac{\Omega_b}{\Omega_M}\right] \left(\frac{D_L^*}{D_L}\right)^{3/2}\,.
\end{equation}
(Refs.~\cite{Allen:2007ue,Allen:2011zs,Ettori:2009wp,Mantz:2014xba,Holanda:2020sqm,SPT:2021vsu} give the details on this technique.) In Eq.~(\ref{fgas1}), $K(z)$ stands for the calibration constant which is equal to $0.96\pm0.12$~\cite{Mantz:2014xba} and $A(z)$ represents the angular correction factor which is close to unity. The asterisk denotes the corresponding quantities for the fiducial model used in the observations to obtain  $f_{gas}$ (usually a flat $\Lambda$CDM model with Hubble constant $H_0=70$ km s$^{-1}$ Mpc$^{-1}$ and the present-day total matter density parameter $\Omega_M=0.3$), finally, $\gamma(z)$ is the depletion parameter, which indicates the amount of cosmic baryons that are thermalized within the cluster potential---see details in Refs.
~\cite{Allen:2007ue,Allen:2011zs,Battaglia:2012th,Planelles:2012vp,Mantz:2014xba}. It is very important to comment that the cosmological analyses with gas mass fraction measurements are model-independent due to the ratio in the parenthesis of Eq.~(\ref{fgas1}). This factor  takes into account the expected variation in the gas mass fraction measurement when the underlying cosmology is modified. We perform our statistical analysis using $\gamma(z)$ as  a constant factor, whose value was obtained from hydrodynamical simulations:\footnote{Recent observational analyses have supported this value, cf. Refs.
~\cite{Holanda:2017cmc,Bora:2021iww}.}  $\gamma_0=0.85 \pm 0.08$ \cite{Battaglia:2012th,Planelles:2012vp}. 

Therefore, for a redshift-dependent speed of light given by $c(z)=c_0\phi_{c}(z)$ and $z$-dependent gravitational coupling $G(z) \propto c^4(z)$, Eq.~(\ref{fgas1}) must be modified according to Eq.~(\ref{final}), i.e.:
\begin{equation}
\label{fgas2}
f_{gas}(z) = \phi_{c}^{11/2}(z) \gamma(z)K(z) A(z) \left[\frac{\Omega_b}{\Omega_M}\right] \left(\frac{D_L^*}{D_L}\right)^{3/2}\,.
\end{equation}
It is possible to constrain $\phi_{c}(z)$ if one knows the luminosity distance to the   galaxy cluster. Here, the luminosity distance for each galaxy cluster of the sample is obtained by using SNe Ia data in  \cite{Pan-STARRS1:2017jku}. Our sample of $D_L$ is constructed from the apparent magnitude ($m_b$) of the Pantheon catalog (Fig.~\ref{fig:riess1}--right) by considering $M_b=-19.23 \pm 0.04$ (the absolute magnitude) through the relation
\begin{equation}
    D_L=10^{(m_b - M_b - 25)/5} \mathrm{Mpc}.
    \label{D_L(mu)}
\end{equation}
However, in order to perform the appropriate tests, we must use SNe Ia at (approximately) the same redshift of the each galaxy cluster. Thus, we make a selection of SNe Ia according to the criterion: $| z_{GC}-z_{SNe}| \leq 0.005$. Then, we perform the weighted average for each measurement utilising:
\begin{equation}
\bar{D}_L = \frac{\sum_i {D_L}_i/\sigma_{{D_L}_i}^{2}}{\sum_i 1/\sigma_{{D_L}_i}^{2}}, 
\label{eq:D_L-bar}
\end{equation}
\begin{equation}
\sigma_{\bar{D}_L}^2 = \frac{1}{\sum_i 1/ \sigma_{{D_L}_i}^{2}}.
\end{equation}
It is worth mentioning that Pantheon \cite{Pan-STARRS1:2017jku} is the most recent wide refined sample of SNe Ia observations found in the literature; it consists of 1049 spectroscopically confirmed SNe Ia and covers a redshift range of $0.01 \leq z \leq 2.3$.

{At this point, it should be emphasized that Eq.~(\ref{eq:D_L-bar}) does not use a luminosity distance where $c=c(z)$. Therein, $\bar{D}_L$ is the the average value weighted by the uncertainties associated with the measurements of the SNe Ia distance modulus $\mu\equiv \left( m_b-M_b \right)$, cf. Eq.~(\ref{D_L(mu)}). Our working hypotheses are that the SNe Ia apparent magnitude data directly determine their luminosity distance and that any eventual correction to $D_L$ coming from $c(z)$ is subdominant with respect to the pre-factor $\phi_c^{11/2}(z)$ already present in Eq.~(\ref{fgas2}).} \footnote{{In actuality, an eventual dependence of $D_L$ on $c(z)$---such as that considered by \cite{Cuzinatto:2022dta}---would contribute to reduce the power of $\phi_c(z)$ in Eq.~(\ref{fgas2}), thus decreasing the effect produced by a hypothesized varying $c$ on $f_{gas}$. This would make even more difficult for us to assess if $c$ is in fact $z$-dependent via $f_{gas}$ data. Our current analysis (disregarding the contribution of $c(z)$ in $D_L$) therefore provides a prediction of the maximum effect of the varying $c$ according to the $f_{gas}$ observational window.}}

Eq.~(\ref{fgas2}) requires the value of the ratio $\Omega_b/\Omega_M$. In order to make our analyses independent of the Planck satellite results, we consider the value of this ratio as obtained by \cite{SPT:2021vsu}: $0.150 \pm 0.021$.  This bound is obtained from six gas mass fraction measurements of galaxy clusters with $z<0.16$ and by using the SH0ES results \cite{Riess:2019cxk} for the Hubble constant. The advantage in doing so is that this bound is insensitive to the particular model of dark energy used, since that its equation of state does not evolve strongly over this redshift range. Then, for further statistical analyses, we exclude these six gas mass fraction data with $z<0.16$, and thereby use only the remaining 38 $f_{gas}$ data (with $z>0.16$ ).


\section{Constraining the minimal CPC model \label{sec-GMFconstrainsCPC}}

The constraints on the model parameter $c_1$ can be obtained from Eq.~(\ref{fgas2}) by maximizing the likelihood function, ${\cal{L}}\propto e^{-\chi^2/2}$, where $\chi^2$ is  given by:
\begin{equation}
    \label{eq:logL2}
  \chi^2 =  \sum_{i=1}^{38}\frac{\left(\phi_c^{11/2}(z)- \left[\frac{f_{gas}}{K(z) A(z) \gamma_0}\right]\left[\frac{\Omega_M}{\Omega_b}\right]\left[\frac{D_L}{D^*_L}\right]^{3/2}\right)^2}{\sigma_{i}^2} .
\end{equation}      
Here, $\sigma_i$ denotes statistical errors associated with the gas mass fraction measurements and the Pantheon sample; they are obtained by using standard error propagation methods.  

We consider the {three} cases for $\phi_{c}(z)$ in Eq.~(\ref{eq:c(z)}): 
\begin{itemize}
\item[(I)] $\quad \phi_{c}(z) = 1+c_1 z$
\item[(II)] $\quad \phi_{c}(z) = 1+c_1 z/(1+z)$
\item[(III)] $\quad \phi_{c}(z) = 1 + c_{1} \log\left(1+z\right)$
\end{itemize}

\begin{figure}
    \centering
    \includegraphics[width=0.8\linewidth]{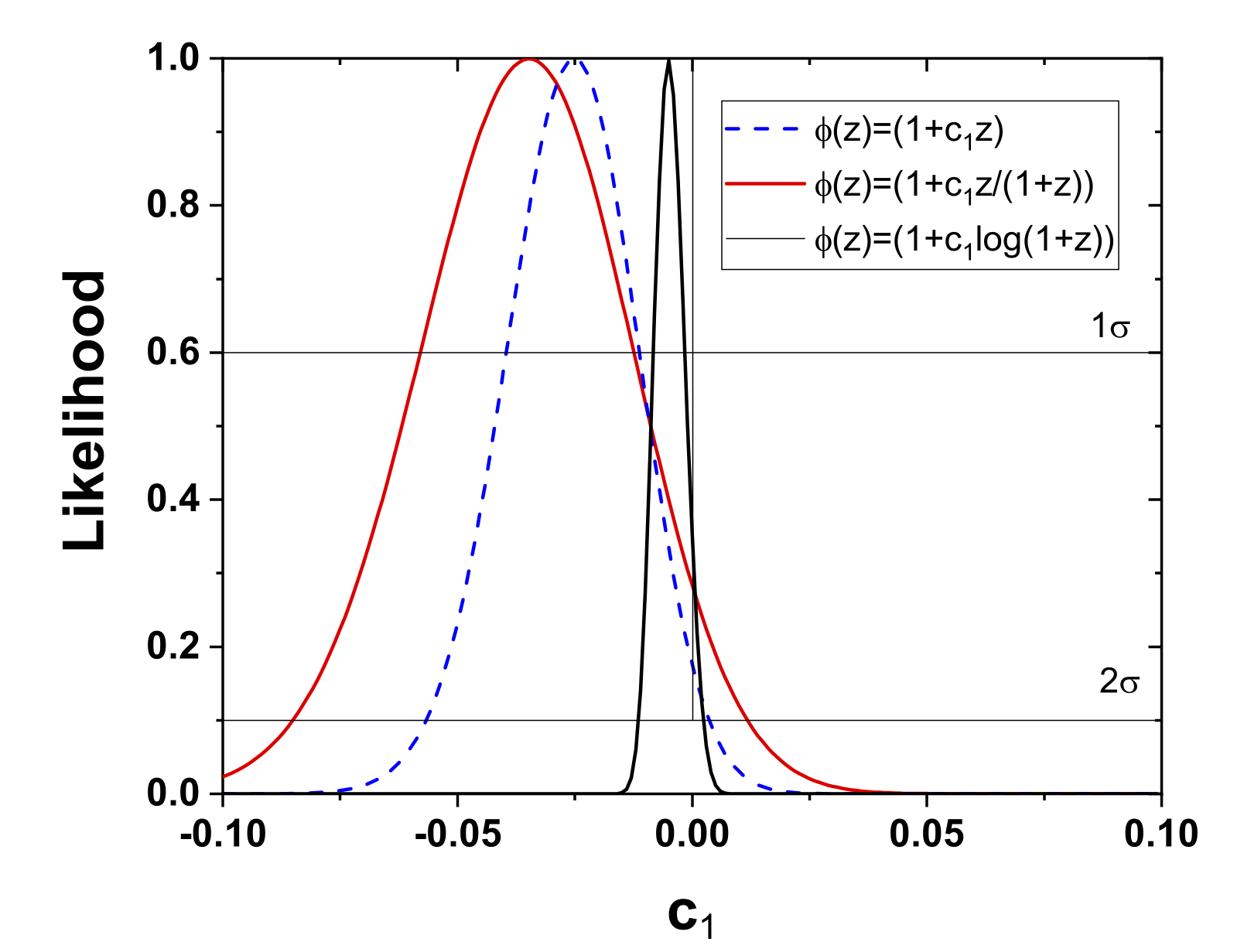}
    \caption{The likelihoods for {three} cases of $\phi_{c}(z)$.}
    \label{fig:riess2}
\end{figure}

We plot the results from our analyses in Fig.~\ref{fig:riess2}. The blue-dashed, red-solid, and black-solid lines correspond to results from the forms (I), (II), and (III) of the function $\phi_{c}(z)$. We obtain (at 1 and 2$\sigma$ c.l.): $c_1=-0.025 \pm 0.012  \pm 0.027$, $c_1=-0.037 \pm 0.025  \pm 0.038$, and $c_1= -0.003 \pm 0.005$ for the parameterizations (I), (II), and (III), respectively. Hence, the standard result ($c_1=0$) is recovered within 2$\sigma$ c.l. for all the three cases.  It is worth comparing our results with those in Ref.~\cite{Mendonca:2021eux}. Therein, the authors considered that only $c$ was varying, with a $\phi_{c}(z)$ function such as that in our case (I); they obtained $c_1=-0.069 \pm 0.056$ and $c_1=-0.198 \pm 0.052$ from analyses involving gas mass fraction plus $H(z)$ data and from gas mass fraction plus SNe Ia observations, respectively. Clearly, the results are in total mutual agreement.


{The negative results regarding a possible variation of $c$---and, consequently, of $G$---occur for the linear parameterizations of $c$ in terms of $z$ and $a$---P (I) and P (II). It is interesting to note that the non-linear parameterization in terms of $\log(z)$---P (III)---also favors a non-varying $c$, and it does so within an even more stringent uncertainly interval than that of P (I) and P (II). Of course, the constancy of the pair $\{c,G\}$ was checked for the particular parameterizations in Eq. (\ref{eq:c(z)}) with the particular data set from galaxy cluster gas mass fraction. Other parameterizations for $c(z)$ and other observational windows could lead to different conclusions.} 

{The $f_{gas}$ data is related to relatively low redshifts. For these redshift values one expects that slowly varying functions for $c(z)$ would fit the data well. In fact, our results support this conjecture. This is true even for parameterization (III) of $c(z) \propto \log(z)$. However, for higher redshifts, one might need more rapidly varying  functions for $c$ in terms of the redshift. Ref.~\cite{Gupta:2020wgz} suggests a parameterization of this kind by adopting $c(a) \propto \exp(a^{\alpha}-1)$ with $\alpha=1.8$. \cite{Gupta:2020wgz} is then able to to resolve the primordial lithium abundance problem. For doing so, the numerical code AlterBBN \cite{Arbey:2018zfh} utilized to predict the Big-Bang Nucleosynthesis (BBN) of light elements was carefully modified for including the effect of varying constants compliant with the exponential type of ansatz adopted for $c(a)$. It is unclear if our slowly changing parameterizations would be as suited for the enterprise of studying BBN as the form in Ref.~\cite{Gupta:2020wgz}: recall that BBN occurred in the very early universe, in the first few minutes after the Big Bang \cite{ellis2012relativistic}. This suggests another line of investigation: to tailor the BBN-analysis code for including our proposals P (I)--(III) and check for the predictions. We deem this task too extensive to be included in the present paper; we intend to pursue the BBN analysis of our model in a separate paper---just like \cite{Gupta:2020wgz} dedicated an entire article to this important problem. The BBN analysis was also the subject of relevant works by \cite{Ichikawa:2002bt,Ichikawa:2004ju,Landau:2004rj,Clara:2020efx,Coc:2006sx,Dmitriev:2003qq,Berengut:2009js,Mori:2019cfo}. Therein, the lithium abundance problem is addressed by allowing variations of some constants (such as the fine structure constant) and/or parameters (e.g. deuterium binding energy, quark masses, etc.). This shows the subject is highly non-trivial.}


\section{Conclusion \label{sec-Conclusion}}

In this paper we have recalled the Co-varying Physical Coupling framework which admits the Einstein Field Equations for gravitational interaction while promoting the couplings $\left\{ c,G,\Lambda\right\} $ to space-time varying functions. Requiring the validity of Bianchi identity and conservation of the stress-energy tensor leads to the General Constraint entangling the simultaneous variation of $\left\{ c,G,\Lambda\right\} $. The simplest model within CPC framework is the one where $\Lambda  =$ constant and $G \propto c^4$. This model was dubbed the minimal CPC model. It was constrained by using  the most recent  $f_{\text{gas}}$ data set: 42 X-ray gas mass fraction measurements of hot ($kT \geq  5$ keV), massive and  relaxed  galaxy clusters from  Chandra archive. The data set consists of galaxy cluster in the redshift range  $0.078 \leq z \leq 1.063$.  

The data analysis constrained the couplings $c=c(z)$ and $G=G(z)$ to be consistent with no variation of these constants, at least for the particular reasonable parameterizations we have used. This conforms with the assumptions in the currently accepted physical theories and it is an important consistency check to their underlying hypotheses. However, it must be emphasised that this conclusion proceeds from a single observational window and should be cross checked with other data sets, such as $H(z)$ cosmic chronometers and type Ia supernovae data. This analysis is relegated to a different paper \cite{Cuzinatto:2022dta} because it is model dependent and relies on the use of the cosmology field equations in the CPC framework. On the other hand, we did not require the CPC analogue of Friedmann equations in this paper. In fact, we did not need to device the luminosity distance expression in the context of CPC framework because any eventual modification to $D_L$ due to the evolution of the co-varying couplings cancels out in the expression $f_{gas}(D_L)$---see Section \ref{sec-GMF}.

Although our observational constraints favour the standard interpretation of $c$ and $G$ as constant couplings, it should be emphasised that more general models within the CPC framework might indicate otherwise. In this direction, we point the reader to the works by \cite{Gupta:2020anq,gupta2021constraint,Gupta:2021wib,Gupta:2020wgz,Gupta:2021eyi,Gupta:2021tma,Gupta:2022amf}, which not only demand $\Lambda$ to vary and adopts more sophisticated ansatze for the time-dependent couplings, but also relax the constancy of other fundamental physical constants, such as $\{\hbar,k_B\}$. The resulting phenomenology is rich and should not be disregarded.

Future perspectives include the study of CPC models including possible variation of $\Lambda$. The solution of the General Constraint for $\Lambda = \Lambda(z)$ is more evolved then the one considered in this paper since it necessarily involves specifying the matter-energy content. This modelling is currently under development. A first step in this direction is Ref.~\cite{Cuzinatto:2022dta}. 

From a observational point of view, it is important to mention that the method discussed here is ready to be used in the near future with the upcoming data set from the X-ray survey eROSITA \cite{eROSITA:2020emt}; this survey is expected to detect $\approx$ 100,000 galaxy clusters.


\section*{Acknowledments}

RRC is grateful to Prof. Rajendra P. Gupta (University of Ottawa) for the hospitality and to CNPq-Brazil (Grant 309984/2020-3) for partial financial support. SHP acknowledges financial support from  {Conselho Nacional de Desenvolvimento Cient\'ifico e Tecnol\'ogico} (CNPq)  (No. 303583/2018-5 and 308469/2021-6.). The authors thank Prof. Rajendra P. Gupta for useful discussions and comments and an anonymous referee whose questions and suggestions lead us to produce a scientifically stronger version of our paper.


\section*{Data Availability}

All data used in this paper are from the references cited.


\section*{Conflict of Interest}

Authors have no conflict of interest.



\end{document}